\documentclass[%
 reprint,
 amsmath,amssymb,
 aps,
]{revtex4-2}

\usepackage{color}
\usepackage{graphicx}
\usepackage{dcolumn}
\usepackage{bm}
\usepackage{hyperref}
\usepackage[all]{hypcap}

\begin{document}

\preprint{APS/123-QED}

\title{Disorder driven crossover between  anomalous Hall regimes in Fe$_3$GaTe$_2$ }

\author{Sang-Eon Lee$^{1,5}$, Minkyu Park$^3$, W. Kice Brown$^4$, Vadym Kulichenko$^1$, Yan Xin$^1$, S. H. Rhim$^6$, Chanyong Hwang$^3$, Jaeyong Kim$^5$, Gregory T. McCandless$^4$, Julia Y. Chan$^4$, Luis Balicas$^{1,2}$}
 \altaffiliation[]{Corresponding author: balicas@magnet.fsu.edu }
\affiliation{%
 $^1$National High Magnetic Field Laboratory, Tallahassee, Florida 32310, USA,\\
 $^2$Department of Physics, Florida State University, Tallahassee, Florida 32310, USA,\\
 $^3$Quantum Technology Institute, Korea Research Institute of Standards and Science, Daejeon, 34113, Republic of Korea,\\
 $^4$Department of Chemistry $\&$ Biochemistry, Baylor University, Waco, Texas 76706, USA,\\
 $^5$Department of Physics, Hanyang University, Seoul 04763, Republic of Korea, \\
 $^6$Department of Semiconductor Physics and Engineering, University of Ulsan, Ulsan 44610, Republic of Korea 
}%

\date{\today}

\begin{abstract}
The large anomalous Hall conductivity (AHC) of the Fe$_3$(Ge,Ga)Te$_2$ compounds has attracted considerable attention. Here, we expose the intrinsic nature of AHC in  Fe$_3$GaTe$_2$ crystals characterized by high conductivities, which show disorder-independent AHC with a pronounced value $\sigma_{xy}^{\text{c}}\approx$ 420 $\Omega^{-1}$cm$^{-1}$. In the low conductivity regime, we observe the scaling relation $\sigma_{xy}\propto\sigma_{xx}^{1.6}$, which crosses over to $\sigma_{xy} \simeq \sigma_{xy}^{\text{c}}$ as $\sigma_{xx}$ increases. Disorder in low-conductivity crystals is confirmed by the broadening of a first-order transition between ferromagnetism and the ferrimagnetic ground state. Through density functional theory (DFT) calculations, we reveal that the dominant sources of Berry curvature are located a few hundred meV below the Fermi energy around the $\Gamma$-point.  Therefore, Fe$_3$GaTe$_2$ clearly exposes the disorder-induced crossover among distinct AHC regimes, previously inferred from measurements on different ferromagnets located in either side of the crossover region.

\end{abstract}

\maketitle


Despite being discovered by Edwin Hall in 1880 \cite{Anomalous_Hall}, the anomalous Hall effect (AHE) continues to be a very active topic of research in condensed matter physics. Based on the seminal works by Karplus $\&$ Luttinger \cite{PhysRev.95.1154, PhysRev.112.739}, Smit \cite{SMIT1955877, SMIT195839}, and Berger \cite{PhysRevB.2.4559}, the modern understanding of the AHE has reached a certain degree of consensus achieved by splitting its behavior according to three conductivity regimes: the dirty regime (for $\sigma_{xx} \lesssim 3\times10^3$ $\Omega^{-1}$cm$^{-1}$), the moderately dirty regime (for $\sigma_{xx} \sim 3\times10^3$ to $5\times10^6$ $\Omega^{-1}$cm$^{-1}$), and the super clean regime (for $\sigma_{xx} \gtrsim 5\times10^6$ $\Omega^{-1}$cm$^{-1}$) \cite{PhysRevB.77.165103}. With the emergence of the concept of Berry curvature \cite{Berry1984QuantalPF, RevModPhys.82.1959} and concomitant topology \cite{PhysRevLett.49.405, PhysRevLett.61.2015, PhysRevLett.95.226801, PhysRevLett.98.106803, PhysRevLett.100.096407, PhysRevLett.108.140405, PhysRevB.83.205101, doi:10.1126/science.1148047, AtopologyDirac, PhysRevLett.103.146401, doi:10.1126/science.1245085, doi:10.1126/science.aaa9297, doi:10.1126/science.1259327, doi:10.1126/science.1234414} in condensed matter physics, the observation of the moderately dirty AHE regime was regarded as the proof of the Karplus-Luttinger or intrinsic mechanism associated with the texture of Berry curvature (topology) within the material \cite{doi:10.1126/science.1234414, doi:10.1126/science.aax8156, Co3Sn2S2, kim2018large}.

\begin{figure*}[ht]
    \centering
    \includegraphics[width=0.95\textwidth]{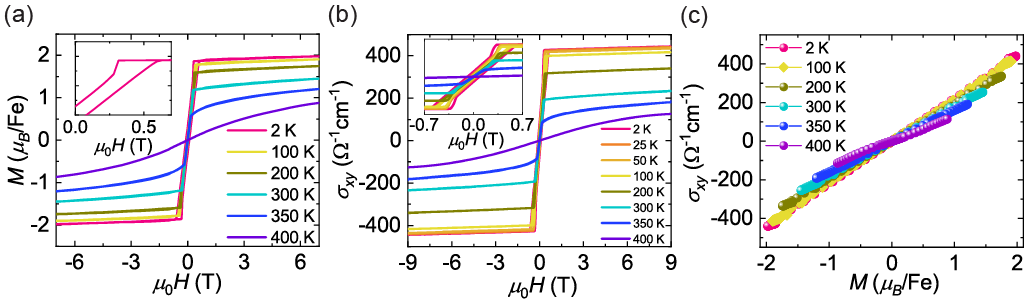}
    \caption{Magnetization and anomalous Hall effect for Fe$_3$GaTe$_2$. (a) Magnetization for Fe$_3$GaTe$_2$ single crystal as a function of the magnetic-field indicating that ferromagnetism is present at temperatures exceeding $T$ = 350 K. Inset: Magnified view of the field-dependent magnetization at $T$ = 2 K within the low field regime, exposing its hysteresis. (b) Hall conductivity as a function of the magnetic field. The ferromagnetic response is clearly observed all the way up to 350 K. Inset: Magnified scale, exposing the hysteresis observed at low fields. (c) $\sigma_{xy}$ as a function of the magnetization at several temperatures. The near proportionality between $\sigma_{xy}$ and $M$ persists to temperatures up to $T$ = 400 K. }
    \label{Fig. 1}
\end{figure*}

Recently, layered ferromagnetic Fe$_3$\textit{X}Te$_2$ compounds, where \textit{X} = Ga or Ge, have attracted a lot of attention because of their distinctive electronic band structure \cite{kim2018large, Cho:2024fa} showing topological character. These compounds also display topological spin textures such as skyrmions \cite{macy2021magnetic, li2024room, zhang2024above, liu2023controllable, https://doi.org/10.1002/adma.202312013} and two-dimensional magnetism beyond room temperature when exfoliated down to a few layers \cite{wang2024hard, zhang2022above}, making them very promising from the fundamental and also application perspectives. In previous studies, avoided band crossings near the Fermi energy, induced by spin-orbit coupling, were claimed to be a major source of Berry curvature and hence of AHE in these compounds \cite{kim2018large, Cho:2024fa}. The large anomalous Hall conductivity (AHC) $\sim$ $e^2/hd$, where $d$ is the spacing between layers, and the observed scaling relation $\sigma_{xy}\propto$ $\sigma_{xx}^{1.6\text{--}1.8}$, usually associated with the dirty regime, was suggested as possible evidence for the intrinsic mechanism \cite{kim2018large, Cho:2024fa}. These studies not only indirectly support the intrinsic mechanism but also imply that the studied samples were slightly outside the moderately dirty regime. However, a fundamental understanding of AHE in these compounds would benefit from a direct observation of the moderately dirty regime. The observation of a moderately dirty regime would also indicate that these materials can reach a nearly disorder-free regime, which is needed to develop applications based on the intrinsic mechanism driven by the Berry curvature. In particular, this is required for compounds like Fe$_3$GaTe$_2$, which displays great potential for spintronics due to its high Curie temperature $T_c \gtrsim 360$ K \cite{zhang2024above}, and the presence of skyrmions above room temperature \cite{li2024room}. 

From the perspective of the fundamental understanding of anomalous Hall effect, clear experimental evidence for the crossover from the dirty to the moderately dirty regime is still lacking. The main challenge is the difficulty in tuning solely the level of disorder in the material while maintaining its chemical composition. Various studies on the scaling behavior of the anomalous Hall effect rely on compositional adjustments to achieve wide ranges of conductivities \cite{lyanda2001charge, matsukura1998transport, edmonds2003magnetoresistance, yuldashev2004anomalous, chiba2007properties, ohno1992magnetotransport, oiwa1999magnetic, iguchi2007scaling, miyasato2007crossover}, which could change the intrinsic electronic and magnetic properties of the studied materials as well as their level of disorder. The investigation on the anomalous Hall effect in Fe thin films evaded this issue by modulating their mobility by adjusting the thickness of the films instead of changing their chemical composition \cite{Proper_scaling}. This study also proposed a scaling analysis, the so-called TYJ scaling, based on the temperature dependence of the conductivity and on the sample-dependent conductivity. Since the dirty regime is not included in the TYJ scale, it is mainly suited to regimes beyond the dirty one above $\sigma_{xx} = 1 \times 10^4$  $ \Omega^{-1}$cm$^{-1}$. Another study claims that a more complete scaling should be applicable even for conductivity regimes beyond $\sigma_{xx} > 1.4 \times 10^5$  $\Omega^{-1}$cm$^{-1}$ \cite{Multivariable_sccaling}. Here, we focus on the lower conductivity regime or when $\sigma_{xx} < 1 \times 10^4$  $ \Omega^{-1}$cm$^{-1}$ and demonstrate a crossover from the dirty to the moderately dirty regimes without tuning the chemical composition, by measuring Fe$_3$GaTe$_2$ crystals, thus complementing our current knowledge of the anomalous Hall effect.

To explore the roles of disorder and Berry curvature texture in Fe$_3$GaTe$_2$, here, we measured 19 single crystals characterized by a wide range of $\sigma_{xx}$ values, from $\sigma_{xx} \simeq 8 \times 10^2$ to $1 \times 10^4$  $\Omega^{-1}$cm$^{-1}$ at $T = 2$ K. Through this wide range of conductivities, we observed a clear dirty regime for $\sigma_{xx}$ below $4 \times 10^3$  $ \Omega^{-1}$cm$^{-1}$, the crossover regime for $\sigma_{xx} = (4 - 7) \times 10^3$  $\Omega^{-1}$cm$^{-1}$, and a saturating regime beyond $\sigma_{xx} = 7 \times 10^3$  $\Omega^{-1}$cm$^{-1}$. Our analysis was performed at the lowest temperature of $T = 2$ K to exclude inelastic effects due to phonons, or magnons. The role of the disorder was confirmed by the observation of sample-dependent broadening of a first-order magnetic phase transition and the behavior of the anomalous Hall coefficient $S_H = \sigma_{xy} / M $ as a function of temperature. Density Functional Theory (DFT) calculations reveal that Fe$_3$GaTe$_2$ displays strong Berry curvature leading to $\sigma_{xy}\approx535$ $\Omega^{-1}$cm$^{-1}$, a value that is slightly higher than the measured one ($\sigma_{xy}\approx420$ $\Omega^{-1}$cm$^{-1}$). DFT calculations also reveal that the dominant sources of Berry curvature are located around the $\Gamma$-point and a few hundred meV below the Fermi energy. This contrasts to  previous studies focusing on the Berry curvature around the $K$-point  located near the Fermi level \cite{kim2018large, Cho:2024fa}.  

Single crystals of Fe$_3$GaTe$_2$ were synthesized through a chemical vapor transport technique. We ordered and labeled all crystals from the highest (C1) to the lowest conductivity (C19). Resistivity $\rho_{xy}$ as a function of $T$ for all samples can be found in Fig.~S1 \cite{supplemental}. Single-crystal X-ray diffraction measurements were performed on four selected single crystals (C2, C6, C9, and C17), whose conductivities ranged from $\sim 2 \times 10^3$ $\Omega^{-1}$cm$^{-1}$ to $\sim 8 \times 10^3$ $\Omega^{-1}$cm$^{-1}$. The precession images and the unit cell are shown in Figs.~S2 and S3, respectively \cite{supplemental} (see also references \cite{SM1, SM2, SM3, SM12, SM13, SM14} therein). In all crystals, we found $\sim 9$ \% of vacancies at the Fe2 site, as well as intercalated Fe atoms at the Fe3 site. We found no correlation between crystal conductivity and its lattice constants. The structural information for these samples is listed in Tables S1 and S2 \cite{supplemental}. High-Angle Annular Dark-Field Scanning Transmission Spectroscopy (HAADF-STEM) imaging reveals positional disorder at the Fe2 and Ga sites with respect to their original position within the mirror plane between  Te atoms, see, Fig. S4 \cite{supplemental} (see also reference \cite{SM4} therein). Intercalated Fe3 ions, fluctuations in Fe2 occupancy, and positional disorder at the Fe2 and Ga sites are the sources of disorder observed in our Fe$_3$GaTe$_2$ single crystals.

\begin{figure*}[ht]
    \centering
    \includegraphics[width=0.9\textwidth]{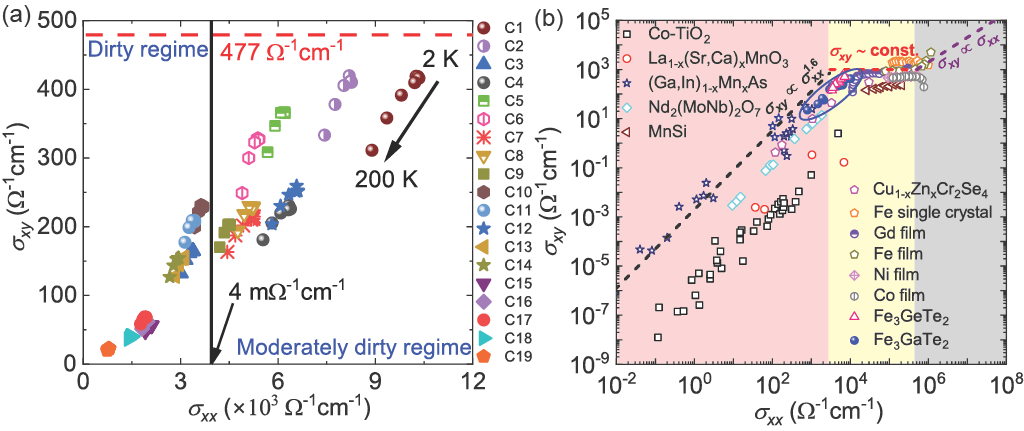}
    \caption{Experimental evidence for a crossover between distinct anomalous Hall regimes in Fe$_3$GaTe$_2$. (a) Hall conductivity $\sigma_{xy}$ as a function of the longitudinal conductivity $\sigma_{xx}$ collected at distinct temperatures on several samples. Values of $\sigma_{xy}$ and $\sigma_{xx}$ were measured under $\mu_0H$ = 1 T and 0 T, respectively. Different symbols are associated with the conductivities of the different samples (i.e., C1 to C19), with their values measured at $T$ = 2, 10, 25, 50, 100, 150, and 200 K. Different scaling regimes are identified below (dirty regime) and above (crossover towards the moderately dirty regimes) the value of $\sigma_{xx}$ = 4  $\times 10^3$  $\Omega^{-1}$cm$^{-1}$. (b) Comparison between the scaling behavior of Fe$_3$GaTe$_2$ and the scaling observed in other compounds as well as with theoretical predictions. Our data from Fe$_3$GaTe$_2$ is indicated by blue markers. Dirty regime ($\sigma_{xy}\propto$ $\sigma_{xx}^{1.6}$), moderately dirty regime ($\sigma_{xy} \sim$ constant), and clean regime ($\sigma_{xy}\propto$ $\sigma_{xx}$) are denoted by red, yellow, and grey shaded areas, respectively. Dashed lines indicate the expected scaling behaviors for each regime. The scaling behavior of Fe$_3$GaTe$_2$ spans both the dirty and moderately dirty regimes (which is dominated by the intrinsic contribution or the Berry curvature). This panel is a modified version of Fig.~12 in Ref.~\cite{PhysRevB.77.165103}, with the data taken from Refs.~\cite{cho2006magnetic, ueno2007anomalous, ramaneti2007anomalous, toyosaki2004anomalous, higgins2004hall} for Co-TiO$_2$; Ref.~\cite{lyanda2001charge} for La$_{1-x}$(Sr,Ca)$_x$MnO$_3$; Refs.~\cite{matsukura1998transport, edmonds2003magnetoresistance, yuldashev2004anomalous, chiba2007properties, ohno1992magnetotransport, oiwa1999magnetic} for (Ga,In)$_{1-x}$Mn$_x$As; Ref.~\cite{iguchi2007scaling} for Nd$_2$(MoNb)$_2$O$_7$; Ref.~\cite{lee2007hidden} for MnSi; Ref.~\cite{miyasato2007crossover} for Cu$_{1-x}$Zn$_x$Cr$_2$Se$_4$, Fe single-crystal, Gd film, Fe film, Ni film, Co film, SrRuO$_3$, and La$_{1-x}$Sr$_x$CoO$_3$; Ref.~\cite{kim2018large} for Fe$_3$GeTe$_2$. }
    \label{Fig. 2*}
\end{figure*}

We measured both the magnetization and the electrical transport properties to analyze the anomalous Hall effect. Figure ~\ref{Fig. 1} displays the magnetization $M(\mu_0H, T)$ and Hall conductivity $\sigma_{xy}(\mu_0H, T)$ as functions of the magnetic field $\mu_0H$, as well as their relation, as observed in crystal C1 for several $T$s. The values of $\sigma_{xy}$ are given by the relation $\sigma_{xy} = \rho_{yx}/(\rho_{xx}^2+\rho_{yx}^2)$. $\rho_{yx}$ and the magnetoresistivity as a function of $\mu_0H$ for all samples can be found in Figs.~S5 and S6, respectively \cite{supplemental}. Both magnetization and $\sigma_{xy}$ show hysteretic and ferromagnetic behavior for $T$s up to 350 K. The hysteresis loops show a step when the sample is brought back from the saturation region, observed in both the magnetization and anomalous Hall conductivity, see insets in Figs.~\ref{Fig. 1}(a,b). This is an indication of an additional magnetic phase transition. When saturated, $\sigma_{xy}$ reaches the large value of $\approx$ 420 $\Omega^{-1}$cm$^{-1}$. This is comparable to previous values of 540 $\Omega^{-1}$cm$^{-1}$ and 680 $\Omega^{-1}$cm$^{-1}$ observed in Fe$_3$GeTe$_2$ \cite{kim2018large} and Fe$_3$GaTe$_2$ \cite{Cho:2024fa}, respectively.  Figure \ref{Fig. 1}(c) shows the plot of $\sigma_{xy}$ as a function $M$ for several \textit{T}s,  exposing the nearly linear dependence $\sigma_{xy}$ on $M$. This indicates that $\sigma_{xy}$ is dominated by the anomalous Hall term, which is proportional to $M$. This proportionality can be expressed through the anomalous Hall coefficient, $S_H$, as $\sigma_{xy} = S_H(\mu_0 H, T) M$. $S_H$ remains nearly constant as a function of the magnetic field at a fixed $T$, but $S_H(\mu_0 H, T)$ gradually decreases with increasing \textit{T}, due to phonon and magnon scattering \cite{PhysRevB.77.014433}. 

\begin{figure*}[ht]
    \centering
    \includegraphics[width=0.9\textwidth]{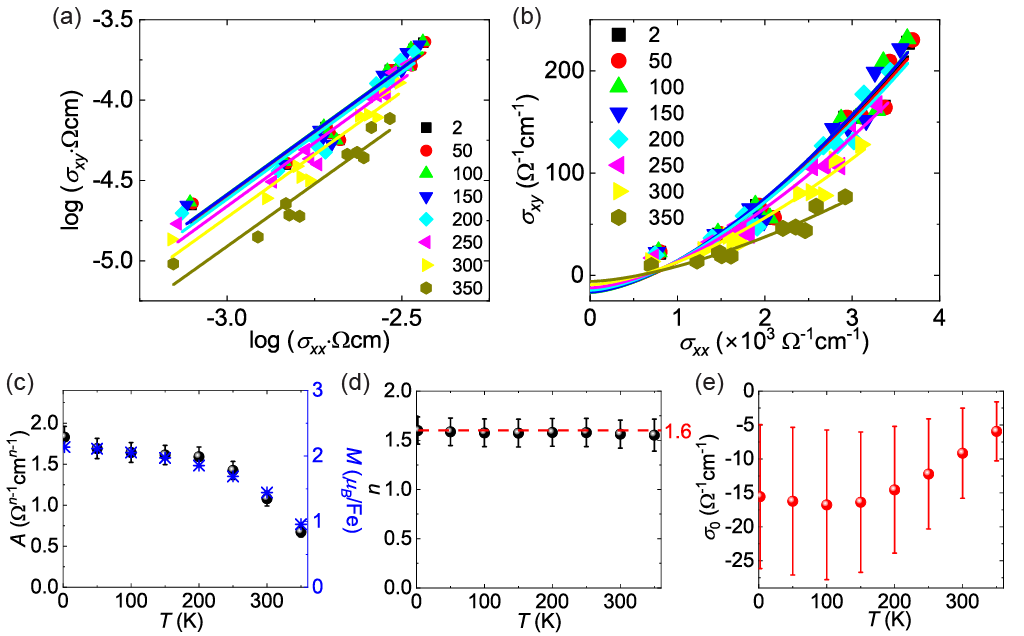}
    \caption{Scaling analysis of the anomalous Hall response of Fe$_3$GaTe$_2$ within the dirty regime. Experimental $\sigma_{xy}$ as a function of $\sigma_{xx}$ is fit to the expression $\sigma_{xy} = A\sigma_{xx}^n + \sigma_0$. (a) Log $\sigma_{xy}$ as a function of log $\sigma_{xx}$ for several temperatures. Solid lines are linear fits whose slopes yield the values of $n$. (b) $\sigma_{xy}$ as function $\sigma_{xx}$ for several temperatures. We fixed the $n$ values to those obtained from the fits in (a) and then fitted the data to $\sigma_{xy} = A\sigma_{xx}^n + \sigma_0$, to extract $A$ and $\sigma_0$. Solid lines represent the fits. (c-d) Resulting $A$, $n$, and $\sigma_0$ values as functions of $T$ from the fittings in (a and b). $A$ values compared to the magnetization under $\mu_0H$ = 1 T from sample C17, displaying agreement in their $T$ dependence. $n$ values are very close to 1.6 regardless of the value of $T$, thus aligning with the theoretical predictions for the dirty regime \cite{PhysRevB.77.165103}. $\sigma_0$ displays large error bars, or a large incertitude in its values.}
    \label{Fig. 3*}

\end{figure*}

We studied the dependence of $\sigma_{xy}$ on the conductivity $\sigma_{xx}$ to understand the underlying mechanism triggering the AHC in Fe$_3$GaTe$_2$. Figure \ref{Fig. 2*}(a) displays $\sigma_{xy}$ as a function of $\sigma_{xx}$ for all 19 samples studied within the temperature range $T = 2$ K - 200 K. $\sigma_{xy}$ and $\sigma_{xx}$ were measured under $\mu_0 H$ = 1 and 0 T, respectively. We identified two distinctive regimes in the $\sigma_{xy}(\sigma_{xx})$ plot, which are separated by $\sigma_{xx} \simeq$ 4 m$\Omega^{-1}$cm$^{-1}$. The $\sigma_{xy}(\sigma_{xx})$ plot indicates that Fe$_3$GaTe$_2$ shows the dirty regime  below $\sigma_{xx} = 4 \times 10^3$  $\Omega^{-1}$cm$^{-1}$ which gradually transits to the saturating, moderately dirty regime as the conductivity increases. This observation implies that the AHC of Fe$_3$GaTe$_2$ is governed by the intrinsic mechanism in the moderately dirty regime due to a strong Berry curvature. The saturating value of $\sigma_{xy}$ in the moderately dirty regime is close to the value of $e^2/hd \simeq 477$ $\Omega^{-1}$cm$^{-1}$, which also points to the intrinsic mechanism \cite{PhysRevLett.97.126602, PhysRevB.77.165103}. In the dirty regime, we found that the relation $\sigma_{xy}\propto\sigma_{xx}^{1.6}$ is satisfied in the temperature range \textit{T} = 2 K to 350 K, which is shown in Fig. \ref{Fig. 3*}, as theoretically predicted and empirically reported \cite{PhysRevLett.97.126602, PhysRevB.77.165103, RevModPhys.82.1539}. 

\begin{figure}[ht]
    \centering
    \includegraphics[width=0.45\textwidth]{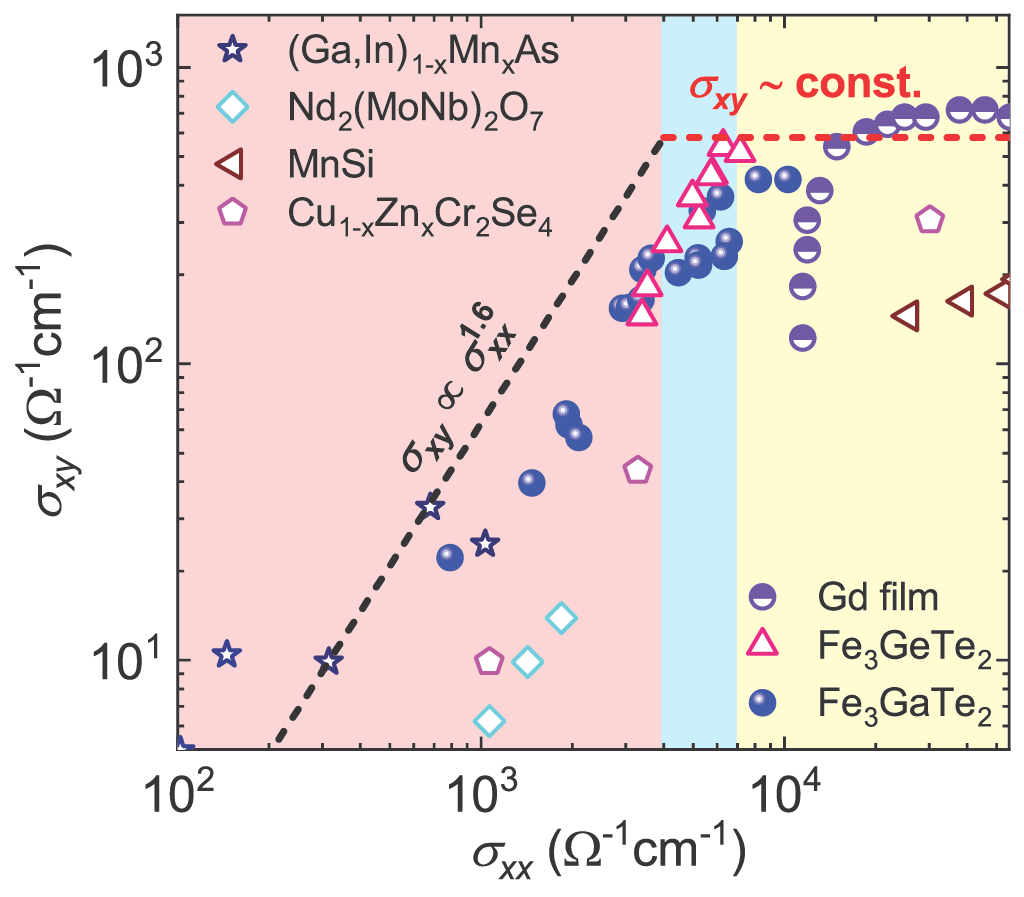}
    \caption{$\sigma_{xy}$ as a function of $\sigma_{xx}$ in a magnified scale focusing on the dirty to moderately dirty and crossover regimes. Red, yellow, and blue shaded areas represent the dirty regime below $\sigma_{xy}$ = 4 $\times 10^3$  $\Omega^{-1}$cm$^{-1}$, the moderately dirty regime above $\sigma_{xy}$ = $7 \times 10^3$  $\Omega^{-1}$cm$^{-1}$, and the intermediate or crossover regime between them, respectively.}
    \label{Fig. 4*}
\end{figure}

Figure \ref{Fig. 2*}(b) displays the relation between $\sigma_{xy}$ and $\sigma_{xx}$ for different compounds, with this panel being a modified version of Fig.~12 in Ref.~\cite{PhysRevB.77.165103}. In that work, the crossover between the dirty and moderately dirty regimes is proposed to be close to the conductivity value $\sigma_{xx} \simeq 3 \times 10^3$  $\Omega^{-1}$cm$^{-1}$, which is consistent with our result of $\sigma_{xx} \simeq 4 \times 10^3$  $\Omega^{-1}$cm$^{-1}$. Figure \ref{Fig. 4*} displays $\sigma_{xy}$ as a function of $\sigma_{xx}$ in a magnified scale that focuses on the data from Fe$_3$GaTe$_2$. We found that Fe$_3$GaTe$_2$ displays the dirty regime behavior below $\sigma_{xx} = 4 \times 10^3$  $\Omega^{-1}$cm$^{-1}$, saturating in the moderately dirty regime above $\sigma_{xx} = 7 \times 10^3$  $\Omega^{-1}$cm$^{-1}$ with the crossover region located between both values. This scaling for Fe$_3$GaTe$_2$ is akin to the one observed for its isomorphic compound Fe$_3$GeTe$_2$ \cite{kim2018large} although Fe$_3$GeTe$_2$ did not clearly expose the moderately dirty regime. Therefore,  Fe$_3$GaTe$_2$ is a rare example of a compound that spans distinct AHC regimes, allowing us to confirm the existence of a crossover beyond a critical conductivity value.

Figure \ref{Fig. 3*} shows the detailed scaling analysis of the dirty regime below $\sigma_{xx} = 4 \times 10^3$  $\Omega^{-1}$cm$^{-1}$. We analyzed the scaling relation by fitting the data to $\sigma_{xy} = A\sigma_{xx}^n + \sigma_0$. First, we took the logarithmic values of both $\sigma_{xy}$ and $\sigma_{xx}$, which are plotted in Fig. \ref{Fig. 3*} (a), to extract the value of $n$. Subsequently, we fit $\sigma_{xy}$ as a function of $\sigma_{xx}$ to the relation $\sigma_{xy} = (A\sigma_{xx}^n + \sigma_0)$ for fixed values of $n$ obtained from the logarithmic plot. These fits yield $A$ and the constant $\sigma_0$ (see Fig. \ref{Fig. 3*} (b)). The dependence on temperature of $A$, $n$, and $\sigma_0$ are plotted in Figs. \ref{Fig. 3*} (c to e), respectively. The factor $A$ just follows the magnetization and is in line with our intuition. The power $n$ is close to the expected value $n \sim$ 1.6 even up to $T$ = 350 K, which is just below the Curie temperature. This clearly shows that Fe$_3$GaTe$_2$ is in the dirty regime below $\sigma_{xx} = 4 \times 10^3$  $\Omega^{-1}$cm$^{-1}$. Finally, we find that the values of $\sigma_0$ are within the error bars, suggesting that the strict scaling relation $\sigma_{xy} = A\sigma_{xx}^n$ is satisfied within the dirty AHC regime of Fe$_3$GaTe$_2$.

\begin{figure}[ht]
    \centering
    \includegraphics[width=0.45\textwidth]{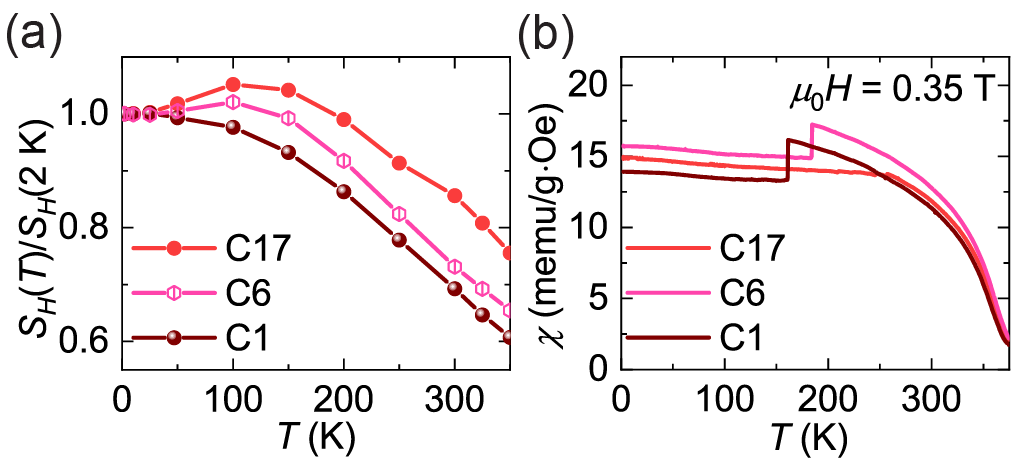}
    \caption{Anomalous Hall coefficient $S_H$ and magnetic susceptibility $\chi$ for samples located within different scaling regimes. Samples C1, C6, and C17 were selected from the moderately dirty regime, the intermediate regime, and the dirty regime, respectively. (a) Anomalous Hall coefficient ($S_H =\sigma_{xy}/M$) for all three crystals as a function of \textit{T} measured under $\mu_0 H$ =  1 T and normalized by the value of $S_H$ collected at $T =$ 2 K. (b) Magnetic susceptibility $\chi$ collected from the three crystals under $\mu_0 H$ =  0.35 T. As the sample quality decreases, the transition temperature tends to increase, becoming non-observable as $\sigma_{xx}$ decreases.}
    \label{Fig. 5*}

\end{figure}

The disorder that affects the value of $\sigma_{xx}$ also affects the temperature dependence of $S_H$. It also considerably broadens the first-order transition observed between ferromagnetism and the ferrimagnetic ground state of Fe$_3$GaTe$_2$. Figure  \ref{Fig. 5*}(a) shows $S_H$ as function of \textit{T} collected under $\mu_0 H$ =  1 T for samples C1, C6, and C17, which are in the moderately dirty regime, at the boundary between the dirty and moderately dirty regimes, and in the dirty regime, respectively. The magnetization data for samples C6, and C17 can be found in Fig.~S7 \cite{supplemental}. For sample C1 located within the moderately dirty regime, $S_{H}$ monotonically decreases with increasing \textit{T}. In contrast, as the crystalline disorder becomes more prominent, a distinct behavior is observed in $S_{H}$ at low temperatures; $S_H$ increases with $T$. In the moderately dirty regime, the AHC signal is robust with respect to elastic scattering, but it becomes considerably suppressed by the inelastic scattering (e.g., spin-flips due to magnetic impurities). This is also illustrated by the rapid decrease of $\sigma_{xy}$ in the moderately dirty regime upon increasing \textit{T}, see Fig.~\ref{Fig. 2*} (a).

The effect of disorder on the magnetic response of Fe$_3$GaTe$_2$ is exposed by the behavior of the first-order magnetic transition. Figure \ref{Fig. 5*} (b) shows the magnetic susceptibility of samples C1, C6, and C17 as a function of \textit{T}. These measurements were taken under field-cooled conditions at $\mu_0 H$ =  0.35 $T$. For the highest Hall conductivity sample C1, the magnetic transition occurs at $T$ $\approx$ 161 K leading to a sharp step that reduces the magnetization by $\sim$ 17.5 \%. For the intermediate quality sample C6, at the boundary between both regimes, the transition temperature is shifted to $T$ $\approx$ 184 K, while the magnetization decreases by just 13.0 \%. The smaller decrease in magnetization is evidence for spatial inhomogeneity leading to the suppression of the first-order transition. For sample C17, located in the dirty regime, the phase transition is completely suppressed due to a more severe structural disorder.

\begin{figure*}[ht]
    \centering
    \includegraphics[width=.7\textwidth]{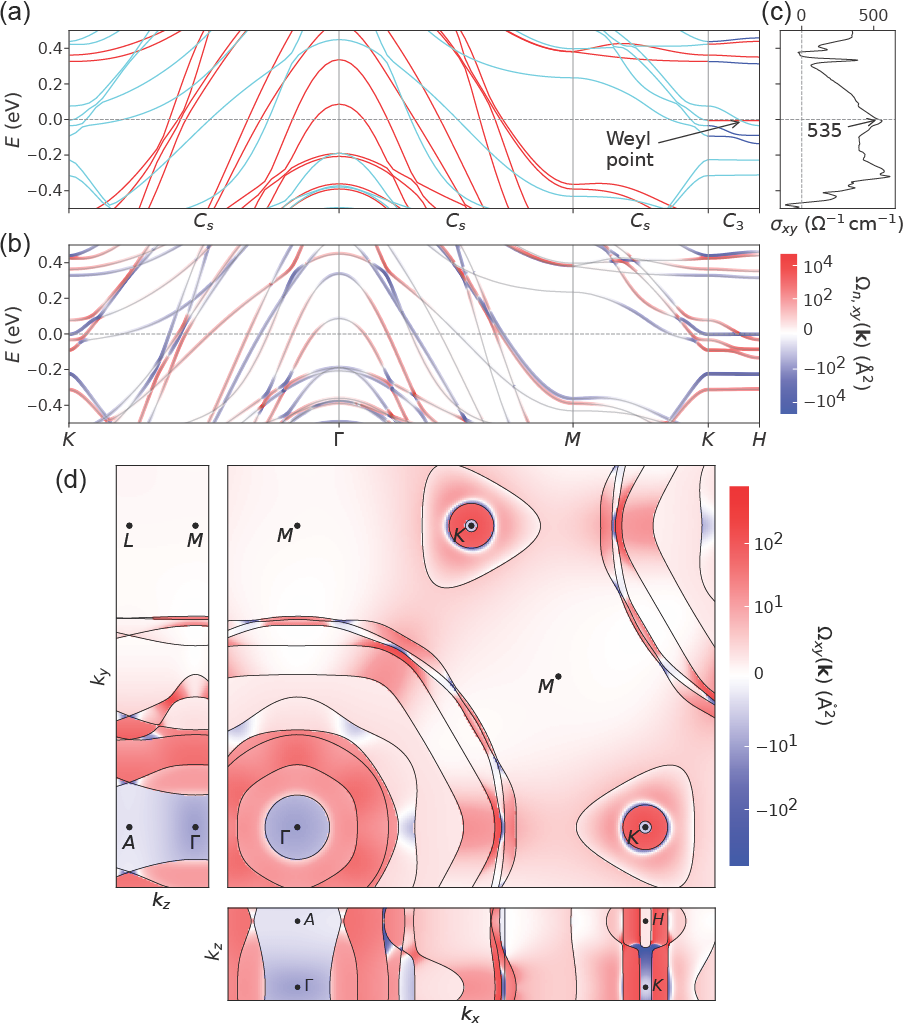}
    \caption{Electronic band structure and distribution of Berry curvature in Fe$_3$GaTe$_2$. (a) Electronic band structure along the high symmetry points within its Brillouin zone. Eigenstates with eigenvalues of $-i$ and $i$ with respect to the $z$-direction mirror operation $C_s$ are depicted by red and cyan lines along the $K$--$\Gamma$, $\Gamma$--$M$, and $M$--$K$ directions. Along the $K$--$H$ direction, the eigenstates with the eigenvalue of $e^{-2\pi i/3}$, $e^{+2\pi i/3}$, and $1$ with respect to the three-fold rotation $C_3$ operation about the $z$-direction are depicted by red, cyan, and blue lines. The band crossing point along the $K$--$H$ path is protected by the three-fold symmetry, forming a Weyl point. (b) Band-resolved Berry curvature along high-symmetry directions. (c) Anomalous Hall conductivity as a function of chemical potential. 
(d) Momentum-resolved Berry curvature map on the three perpendicular $k$-planes. Main Berry curvature contribution is found around the $\Gamma$-- and $K$--points. 
Note the strong negative contributions along the $K$--$H$ path. 
}
    \label{Fig. 6*}
\end{figure*}

To understand the origin of the strong Berry curvature contributing to the intrinsic anomalous Hall effect mechanism in Fe$_3$GaTe$_2$, we conducted DFT calculations (see Supplemental Material and references \cite{SM5, SM6, SM7, SM8, SM9, SM10, SM11} therein for the details). Figure \ref{Fig. 6*} (a) shows the calculated electronic band structure of Fe$_3$GaTe$_2$ along the high-symmetry directions within its Brillouin zone. 
The states along the $K$--$\Gamma$, $\Gamma$--$M$, and $M$--$K$ directions can be classified by the eigenvalues of the mirror operator $C_s$ around the crystallographic $c$-axis. 
Likewise, the states on the $K$--$H$ path can be classified by the eigenvalues of the three-fold $C_3$ rotation operation around the crystal $c$-axis. 
Near the $K$--point, the Fermi energy crosses the band edges, leading to small Fermi surface (FS) pockets. We observe a Weyl node formed along the $K$--$H$ direction, which is protected by the band crossing between states having distinct eigenvalues. 
This node is the trace of a nodal line that is protected by the $C_{3v}$ symmetry in the absence of spin-orbit interaction. Large FS pockets are formed near the $\Gamma$--point.

Band resolved Berry curvature is shown in Fig.~\ref{Fig. 6*} (b). 
The Berry curvature is distributed throughout all regions of the Brillouin zone. We identified the energy regimes where the Berry curvatures are particularly strong by calculating the chemical potential-dependent value of the Hall conductivity $\sigma_{xy}$, which is shown in Fig.~\ref{Fig. 6*} (c). 
The momentum-resolved Berry curvature map indicates that the Berry curvature contribution is particularly strong around the $\Gamma$-- and $K$-- points, as shown in Fig.~\ref{Fig. 6*}(d). The reader can compare this panel with the distribution of Berry curvature,  at different values of the chemical potential, displayed in Fig.~S8  \cite{supplemental}.
Around the $\Gamma$-point, the distribution of Berry curvature is consistently similar regardless of the position of the chemical potential. 
On the other hand, around the $K$--point, the distribution and sign of the Berry curvature become strongly dependent on the chemical potential. 
We conclude that the dependence of the anomalous Hall conductivity as a function of the chemical potential is mainly due to the Berry curvature around the $K$-point, with the $\Gamma$--point providing an overall positive contribution. 

In a simple model, the large contribution of Berry curvature is explained by the specific location of the Fermi level, which is located within the gap or at the band edges between the conduction and valence bands \cite{PhysRevLett.97.126602, PhysRevB.77.165103}. This results from the two-band model, where the contribution of the Berry curvature associated to the conduction and valence bands nearly cancel each other. The Berry curvature distribution near the $K$-point of Fe$_3$GaTe$_2$ behaves similarly. Near the $K$--point, one observes that some bands are paired by providing positive and negative contributions. This is expected since the origin of the Berry curvature along the $K$--$H$ path is a gapped nodal line.  The caveat with this simple model is the resonant increase of the AHC, which requires a precise Fermi-level placement within a narrow window of energies. However, if the Berry curvature distribution results from many bands, i.e. more than two, it can spread over a broad range of energies, which is precisely the result of our calculations around the $\Gamma$--point. 

In this study, we observed a clear crossover between the dirty to the moderately dirty regime in the anomalous Hall response of Fe$_3$GaTe$_2$. The large value of the saturating anomalous Hall conductivity  $\simeq$ 420 $\Omega^{-1}$cm$^{-1}$, which is close to quantized value $\sim$ $e^2/hd \simeq$ 477 $\Omega^{-1}$cm$^{-1}$, coupled to its disorder independence, exposes a significant intrinsic contribution to the anomalous Hall response of Fe$_3$GaTe$_2$. Disorder in samples located in the low-conductivity region, where $\sigma_{xy} \propto \sigma_{xx}^{1.6}$, is supported by a TEM study and the broadening of the first-order transition towards the ferrimagnetic ground state. Therefore, Fe$_3$GaTe$_2$ offers, in a single compound, a clear example of the crossover between dirty and intrinsic regimes, which was inferred from measurements in distinct ferromagnets \cite{miyasato2007crossover, PhysRevLett.97.126602,PhysRevB.77.165103}. Our calculations and analysis reveal that a sizeable Berry curvature distribution around the $\Gamma$--point provides the dominant contribution to the intrinsic anomalous Hall. They also emphasize the importance of the large Fermi surface pockets around the $\Gamma$--point. This contrasts to previous reports that emphasized the importance of the small FS pockets around the $K$--point \cite{kim2018large}, thus deepening our understanding of the intrinsic anomalous Hall regime of Fe$_3$GaTe$_2$.

\begin{acknowledgments}
L.B. acknowledges support from the US DoE, BES program through award DE-SC0002613 (synthesis and measurements), NSF-DMR 2219003 (heterostructure fabrication) and the Office Naval Research DURIP Grant 11997003 (stacking under inert conditions). JC is supported by the US Department of Energy through the DE-SC0022854 award and the Welch Foundation through AA-2056-20240404. The National High Magnetic Field Laboratory is supported by the US-NSF Cooperative agreement Grant DMR-2128556, and the state of Florida. Jaeyong Kim acknowledges support from the National Research Foundation of Korea (NRF) grant funded by the Korean government (MSTI) (No. 2022H1D3A3A01077468). Chanyong Hwang acknowledges support from the Nano $\&$ Material Technology Development Program through the National Research Foundation of Korea (NRF) funded by the Ministry of Science and ICT (RS-2024-00451261). S. H. Rhim acknowledges support from the National Research Foundation of Korea (NRF) grant funded by the Korea government (MSIT) (No. NRF-RS-2024-00449996). S.-E. L. was partially supported by the National Research Foundation of Korea (NRF) grant (Nos. RS-2024-00412446).
\end{acknowledgments}

\bibliography{references}

\end{document}